\documentclass[11pt, a4paper]{article}

\usepackage{jheppub}
\usepackage{mathtools, amsfonts, amsthm, latexsym, amssymb, dsfont, mathrsfs}
\usepackage[T1]{fontenc}
\usepackage{newtxtext}
\usepackage{physics2}
\usephysicsmodule{ab}
\usepackage{braket}
\usepackage{color}
\usepackage{bm}
\usepackage{comment}

\usepackage{tikz}
\usetikzlibrary{decorations.markings}

\usepackage{hyperref}
\hypersetup{colorlinks=true, citecolor=blue, urlcolor=blue}
\usepackage{cleveref}

\usepackage{titlesec}
\titlespacing{\section}
    {0pt}
    {5pt}
    {3pt}
\titlespacing{\subsection}
    {0pt}
    {5pt}
    {3pt}

\AtBeginDocument{
  \abovedisplayskip     =0.7\abovedisplayskip
  \abovedisplayshortskip=0.7\abovedisplayshortskip
  \belowdisplayskip     =0.7\belowdisplayskip
  \belowdisplayshortskip=0.7\belowdisplayshortskip}


\newcommand{\del}{\partial}
\newcommand{\nn}{\nonumber\\}

\begin{document}

\title{Dyonic edge modes in Abelian gauge theory}

\author[a]{Keito Shimizu}
\author[a, b, c]{and Sotaro Sugishita}

\affiliation[a]{Department of Physics, Kyoto University, Kitashirakawa-Oiwakecho, Kyoto 606-8502, Japan}
\affiliation[b]{Department of Physics, Hokkaido University, Sapporo 060-0810, Japan}
\affiliation[c]{RIKEN Center for Interdisciplinary Theoretical and Mathematical Sciences (iTHEMS), RIKEN, Wako 351-0198, Japan}
\date{}

\usetikzlibrary{calc} 

\emailAdd{kate@gauge.scphys.kyoto-u.ac.jp, sugishita(at)particle.sci.hokudai.ac.jp}

\abstract{
    We find new boundary conditions in four-dimensional Abelian gauge theory with general Chern--Simons boundary couplings. The boundary conditions allow dyonic edge modes and physical boundary symmetries.
    In particular, one of the new boundary conditions makes both electric and magnetic charges physical. We also study how our boundary conditions and charges transform under the $\mathrm{SL}(2,\mathbb Z)$ duality.
    }

\preprint{%
\begin{tabular}[t]{@{}r@{}}
KUNS-3128\\
EPHOU-26-12\\
RIKEN-iTHEMS-Report-26
\end{tabular}%
}

\maketitle
\newpage

\section{Introduction}
While gauge transformations are a defining feature of gauge theories, they are usually regarded as redundancies of description.
In the presence of a boundary, however, this interpretation becomes subtle. Pure-gauge components \textit{always} localize on the boundary and can be dynamical degrees of freedom, known as edge modes.
The importance of boundary degrees of freedom or edge modes in gauge theories and gravity has long been recognized \cite{Regge:1974zd, Gervais:1976ec, Benguria:1976in, Wadia:1976fa, Wadia:1977qr}. 
Edge modes have long played an important role in physics, from the fractional quantum Hall effect \cite{Wen:1990se, PhysRevLett.64.216, MacDonald:1990zz, Balachandran:1991dw, Wen:1995qn} to black hole physics \cite{Brown:1986nw, Carlip:1994gy, Balachandran:1994up, Balachandran:1995qa, Ashtekar:1997yu, Carlip:2005zn}.
Moreover, edge modes are closely associated with infrared structures of gauge theory such as memory effects, soft theorems, dressed states and confinement \cite{Balachandran:2013wsa, Strominger:2013lka, Strominger:2013jfa, He:2014laa, Cachazo:2014fwa, Campiglia:2014yka, He:2014cra, Kapec:2014opa, Lysov:2014csa, Liu:2014vva, Strominger:2014pwa, Kapec:2014zla, Larkoski:2014bxa, Kapec:2015vwa, He:2015zea, Campiglia:2015qka, Kapec:2015ena, Campiglia:2015kxa, Pasterski:2015tva, Dumitrescu:2015fej, Strominger:2015bla, Campiglia:2016hvg, Mirbabayi:2016axw, Gabai:2016kuf, Campiglia:2017dpg, Kapec:2017tkm, Hamada:2017atr, Pate:2017fgt, Hamada:2018cjj, Carney:2018ygh, Hirai:2018ijc, 
Hosseinzadeh:2018dkh, Campiglia:2018see, Francia:2018jtb, Neuenfeld:2018fdw, Henneaux:2018mgn, Hirai:2019gio, Gonzo:2019fai, Choi:2019rlz, Choi:2019sjs, Henneaux:2020nxi, Hirai:2020kzx, Campiglia:2021oqz, Hirai:2022yqw, Nagy:2022xxs, Peraza:2023ivy, Nagy:2024jua, Oertel:2026wsm, Oertel:2026oqv,Shimizu:2025hfl}.
In this context, the appearance of dynamical edge modes is closely related to the fact that gauge transformations acting nontrivially at the boundary can become physical symmetries rather than redundancies.

The choice of boundary condition determines whether edge modes are dynamical and whether large gauge transformations act as physical symmetries.
Abelian gauge theory provides a useful testing ground for exploring this dependence. Maxwell theory, in particular, has been studied extensively.
As discussed in \cite{Carrozza:2021gju, Ball:2024hqe, Araujo-Regado:2024dpr, Hoehn:2025pmx, Shimizu:2026zug}, the standard Neumann, Dirichlet, and Robin boundary conditions do not allow edge modes to be dynamical.
In order to make them dynamical, one must impose other boundary conditions such as the dynamical edge mode boundary condition \cite{Ball:2024hqe} or soft boundary conditions \cite{Araujo-Regado:2024dpr}.
These boundary conditions make electric edge modes dynamical and electric charges physical.
In \cite{Shimizu:2026zug}, the authors instead used the electromagnetic duality to identify boundary conditions under which magnetic edge modes become dynamical and magnetic charges become physical.
To our knowledge, no previously proposed boundary condition makes both electric and magnetic charges physical.
Such boundary conditions are necessary in order to discuss the central charge that appears in the commutator between electric and magnetic charges, as considered in \cite{Hosseinzadeh:2018dkh, Freidel:2018fsk}. If either the electric or magnetic charge is not physical but instead generates a gauge redundancy, then the corresponding commutator vanishes on the physical phase space, and so does the associated central term.
It is therefore natural to ask whether there is a boundary condition under which both electric and magnetic charges are physical.

In this paper, we extend the method of electromagnetic duality in \cite{Shimizu:2026zug} to the full $\mathrm{SL}(2,\mathbb Z)$ duality in Abelian gauge theories by allowing general topological interactions on the boundary. 
The generalized boundary terms allow us to classify a broad range of boundary conditions and identify new edge modes, which can be interpreted as endpoints of generic Wilson-'t Hooft lines.
In addition, we find a boundary condition under which the electric and magnetic charges are simultaneously physical.
Thus, the resulting charge algebra provides a setting in which a possible central extension of the algebra of these charges \cite{Hosseinzadeh:2018dkh, Freidel:2018fsk} can be discussed.
Still, we need singular transformations to obtain a nonzero central charge, and thus its physical significance remains ambiguous.

The rest of this paper is organized as follows.
In \cref{sec:review}, we give a brief review of $\mathrm{SL}(2,\mathbb Z)$ transformations of Abelian gauge theory.
We introduce boundary topological quantum field theory (TQFT) couplings to describe generalized Abelian gauge theory in the presence of a boundary and explain how $\mathrm{SL}(2,\mathbb Z)$ transformations act on the boundary.
In \cref{sec:edge mode}, we classify boundary conditions of the generalized Abelian gauge theory introduced in the previous section.
We find new boundary conditions allowing new physical edge modes associated with physical boundary symmetries. 
In particular, we show a new boundary condition under which the electric and magnetic charges are physical simultaneously.
In \cref{sec:transf laws}, we study how duality wall operators implementing $\mathrm{SL}(2,\mathbb Z)$ act on the new boundary conditions and new edge modes.
We confirm that their action reproduces the expected $\mathrm{SL}(2,\mathbb Z)$ transformations.

\section{Review of \texorpdfstring{$\mathrm{SL}(2,\mathbb Z)$}{SL(2,Z)} in Abelian gauge theory}\label{sec:review}
First of all, we briefly review $\mathrm{SL}(2,\mathbb Z)$ duality in Abelian gauge theory.
In this paper, we focus on theories in a (3+1)-dimensional flat spacetime $\mathcal{M}$ with a (2+1)-dimensional boundary $\Delta$, that is the same setup as \cite{Shimizu:2026zug}.
As in \cite{Shimizu:2026zug}, for concreteness we take $\mathcal{M}$ to be a solid cylinder with boundary $\Delta$ at $r=R$, although the specific shape of the boundary is not important for the following discussion.
The bulk metric is $ds^2=-dt^2+dr^2 + r^2\gamma_{ab}d\Omega^ad\Omega^b$, where $\Omega^a$ $(a=1,2)$ are angular coordinates and $\gamma_{ab}$ is the metric on the unit two-sphere $S^2$.
The volume forms on $\mathcal{M}$ and $\Delta$ are given by $\mathrm{vol}_{\mathcal{M}}=\frac{1}{2}\varepsilon_{ab}\sqrt{-g}~dt\wedge dr\wedge d\Omega^a\wedge d\Omega^b$ and $\mathrm{vol}_{\Delta}=\frac{1}{2}\varepsilon_{ab}R^2\sqrt{\gamma}~dt\wedge d\Omega^a\wedge d\Omega^b$, respectively, where $\varepsilon_{ab}$ is the antisymmetric tensor with normalization $\varepsilon_{12}=-\varepsilon_{21}=1$.
These volume forms fix the orientations of the bulk $\mathcal{M}$ and the boundary $\Delta$.

The bulk action of the Abelian theories is given by
\begin{align}
    S = -\frac{1}{2e^2}\int_{\mathcal{M}}F\wedge*F + \frac{\theta}{8\pi^2}\int_{\mathcal{M}}F\wedge F,
\end{align}
where the second term is the theta term, or the Pontryagin term.
On a compact (spin) manifold, the theory is periodic under $\theta\to\theta+2\pi$. In the presence of a boundary, however, $\theta$ need not be periodic.
It is nevertheless convenient to combine $e$ and $\theta$ into the complex coupling
\begin{align}
    \tau = \frac{\theta}{2\pi} + \frac{2\pi i}{e^2}.
\end{align}
We introduce the standard generators $S$ and $T$ of $\mathrm{SL}(2,\mathbb Z)$, acting on the complex coupling as
\begin{align}
    S: \tau\to-\frac{1}{\tau},\quad T: \tau\to\tau+1.
\end{align}
For the field strength $F=dA$, we define the dual field strength by
\begin{align}\label{def:G}
    G := \frac{2\pi}{e^2}\ast F
    -\frac{\theta}{2\pi}F,
\end{align}
and the bulk EoM is given by $dG=0$.

In the presence of a boundary, an $\mathrm{SL}(2,\mathbb Z)$ transformation of a boundary condition can be implemented by fusing a duality wall with the boundary \cite{Gaiotto:2008ak, Kapustin:2009av}. Since any $\mathrm{SL}(2,\mathbb Z)$ transformation can be generated by successive $S$ and $T$ transformations, we particularly consider the $S$ wall and the $T$-wall, which implement the $S$ and $T$ transformations respectively.
The $S$ wall is a codimension-one interface $W$ separating two bulk gauge theories with gauge fields $A$ and $\hat{A}$, whose couplings are related by $\hat{\tau}=-1/\tau$, and has the wall action
\begin{align}\label{eq:S-wall}
    \pm \frac{1}{2\pi}\int_W A\wedge d\hat{A},
\end{align}
where the sign depends on the orientation of the wall $W$. 
When the wall is fused with the boundary $\Delta$, $\hat{A}$ becomes a boundary gauge field on $\Delta$.
The $T$ transformation is implemented by a $T$ wall separating two gauge theories with couplings $\tau$ and $\tau+1$. Its action is
\begin{align}\label{eq:T-wall}
    \pm\frac{1}{4\pi}\int_W A\wedge dA.
\end{align}

In this paper, instead of 
\eqref{eq:S-wall} and \eqref{eq:T-wall}, we introduce dressed wall actions 
\begin{align}
        \pm\frac{1}{2\pi}\int_W a\wedge d\hat{a},\qquad 
        \pm\frac{1}{4\pi}\int_W a\wedge da,
\end{align}
where $a, \hat{a}$ are dressed gauge fields defined as $a=A|_W+d\phi$, $\hat{a}=\hat{A}|_W+d\hat{\phi}$ and
$\phi$, $\hat{\phi}$ are (compact) scalar fields on $W$. 
Such edge modes $\phi$ and $\hat{\phi}$ are introduced to construct a gauge-invariant phase space for a subsystem (see, e.g., \cite{Donnelly:2016auv}).
Edge modes are useful because we can distinguish gauge redundancies from genuine physical symmetries in the presence of a boundary. Gauge transformations act simultaneously on the bulk and edge fields, while independent transformations of the edge field can be discussed separately as possible physical symmetries.
Under gauge transformations, the fields $(A, \phi)$ transform as $(A, \phi) \to (A+d\alpha, \phi-\alpha)$ and thus the dressed field $a$ is gauge invariant.
On the other hand, we can consider a local shift transformation $\phi \to \phi +\alpha$ with $A$ fixed, and this transformation can be a physical symmetry for appropriate boundary conditions.

Motivated by these wall constructions, we now introduce general boundary topological actions that are compatible with $S$ and $T$ transformations as studied in \cite{Witten:2003ya, Gaiotto:2008ak, Kapustin:2009av}.
Following these works, we consider the following action with general boundary terms: 
\begin{align}\label{eq:general action}
    S &= -\frac{1}{2e^2}\int_{\mathcal{M}}F\wedge*F + \frac{\theta}{8\pi^2}\int_{\mathcal{M}}F\wedge F \notag \\
    &\qquad+ \sum_{j,l=1}^n\frac{k_{jl}}{4\pi}\int_{\Delta}a^j\wedge da^l + \frac{1}{2\pi}\int_{\Delta}a\wedge \sum_{l=1}^nv^lda^l + \frac{p}{4\pi}\int_\Delta a\wedge da,
\end{align}
where $a$ and $a^j$ are the gauge-invariant dressed gauge fields defined by $a\equiv A|_\Delta+d\phi$ and $a^j\equiv A^j+d\phi^j$.
The symmetric matrix $[k]_{jl}=k_{jl}=k_{lj}$ and the vector $[\bm{v}]_l=v_l$ are integer valued. 
We assume that $\ker \bm{v}^\mathsf{T}\,\cap\,\ker k = 0$ as argued in \cite{Kapustin:2009av}. 
We also suppose $p$ is an integer.

Defining $a^0\equiv a$ and a $(n+1)\times(n+1)$ matrix
\begin{align}
    K =
    \begin{pmatrix}
        p & \bm{v}^{\mathsf{T}} \\
        \bm{v} & k
    \end{pmatrix}, \qquad \bm{v}^{\mathsf{T}}=(v^1, \dots, v^n), 
\end{align}
one can write boundary terms in the second line of \eqref{eq:general action} as
\begin{align}
    \frac{1}{4\pi}\int_\Delta (a,Kda),
\end{align}
where $(\cdot,\cdot)$ is a canonical inner product for the label $J=\{0,j\}_{1\leq j\leq n}$.
In this notation, under the $S,\ T$ transformations, the matrix changes as
\begin{align}\label{eq:duality-K}
    S:
    K=\begin{pmatrix}
        p & \bm{v}^{\mathsf{T}} \\
        \bm{v} & k
    \end{pmatrix}
    \to
    \begin{pmatrix}
        0 & \begin{pmatrix} 1 &\cdots&0 \end{pmatrix} \\
        \begin{pmatrix} 1 &\cdots&0 \end{pmatrix}^{\mathsf{T}} & K
    \end{pmatrix},\quad
    T:
    \begin{pmatrix}
        p & \bm{v}^{\mathsf{T}} \\
        \bm{v} & k
    \end{pmatrix}
    \to
    \begin{pmatrix}
        p+1 & \bm{v}^{\mathsf{T}} \\
        \bm{v} & k
    \end{pmatrix}.
\end{align}
Note that the size of the matrix $K$ changes under the $S$ transformation because an $S$ wall introduces additional degrees of freedom on the boundary.

\section{Boundary conditions and edge modes in Abelian gauge theory}\label{sec:edge mode}

So far, we have not specified any boundary condition.
To define the theory, we have to impose boundary conditions.
The previous paper \cite{Shimizu:2026zug} considers the case with $\theta=0$ and no boundary action ($K=0$), as well as the case $K=\begin{pmatrix}
        0 & 1 \\
        1 & 0
    \end{pmatrix}$ 
obtained by applying a single $S$ transformation to the $K=0$ case. It is shown that, for suitable choices of boundary conditions, edge modes become physical degrees of freedom rather than gauge redundancies. 
The purpose of this paper is to classify, for the more general action \eqref{eq:general action}, boundary conditions under which physical edge modes arise, and to investigate how these boundary conditions transform under $\mathrm{SL}(2,\mathbb Z)$ transformations.

The variational principle requires the boundary term in the variation of the total action \eqref{eq:general action} to be a total derivative on $\Delta$ \cite{Harlow:2019yfa}.
We restrict attention to the case in which the variation with respect to each field satisfies the following condition:
\begin{align}\label{eq:dc}
    \frac{1}{2\pi}\delta a\wedge \ab(\frac{2\pi}{e^2}*F - \frac{\theta}{2\pi}F + \sum_{l=1}^nv_lda^l + p\,da) &\overset{\Delta}{=} dc, \\
    \label{eq:dcj}
    \frac{1}{2\pi}\delta a^j\wedge\ab(\sum_{l=1}^n k_{jl}da^l + v_j da) &\overset{\Delta}{=} dc^j.
\end{align}
For later convenience, we define
\begin{align}
    B_0 &= \frac{2\pi}{e^2}*F - \frac{\theta}{2\pi}F + \sum_{l=1}^nv_lda^l + p\,da =G +\sum_{l=1}^nv_lda^l + p\,da,
    \\
    B_j &= \sum_{l=1}^n k_{jl}da^l + v_j da.
\end{align}
Using $K$, these two-forms can be written as
\begin{align}\label{eq:B-matrix}
    B_J=\delta_{J0}\, G
    +\sum_{L=0}^n K_{JL}da^L\quad (J=0,1,\ldots,n),
\end{align}
where $\delta_{J0}$ is the Kronecker delta.
The above condition then takes the form\footnote{The variational principle requires only a weaker condition
\begin{align}
    \frac{1}{2\pi} \sum_J\delta a^J\wedge B_J\overset{\Delta}{=}dc.
\end{align}
In \eqref{eq:dc} and \eqref{eq:dcj}, we impose the stronger requirement that the contribution from the variation of each field be a total derivative independently.
}
\begin{align}
\label{eq:requirement}
    \frac{1}{2\pi}\delta a^J\wedge B_J\overset{\Delta}{=}dc^J\quad (\text{for each}\ J=0,1,\cdots,n).
\end{align}
Using this boundary condition and the bulk equation of motion $d*F=0$, we obtain the following pre-symplectic form:
\begin{align}
    \Omega &= \frac{1}{2\pi}\int_{\Sigma} \delta A\wedge\delta G \notag\\
    &\quad - \frac{1}{4\pi}\int_{\del\Sigma}(\delta a,K\delta a) + \frac{1}{2\pi}\int_{\del\Sigma}\delta\phi\cdot\delta G + \int_{\del\Sigma} \sum_{J}\delta c^J,
\end{align}
where $\Sigma$ is a Cauchy slice.
For simplicity, suppose that $c^J=0$. The theory then has the following gauge redundancies: $(A,\phi,A_1,\phi_1,\cdots,A_n,\phi_n)\to(A+d\alpha,\phi-\alpha,A_1+d\alpha^1,\phi_1-\alpha^1,\cdots,A_n+d\alpha^n,\phi_n-\alpha^n)$.
In addition, the action \eqref{eq:general action} seems to have ``local boundary symmetries'' $\phi^J  \to \phi^J +\alpha^J$ $(J=0,1, \dots, n)$.
As argued in \cite{Ball:2024hqe, Araujo-Regado:2024dpr, Shimizu:2026zug}, it depends on the choice of boundary conditions whether these shift symmetries are genuinely physical or gauge redundant.

For example, one may impose the boundary conditions 
\begin{align}\label{allB=0}
    B_J \overset{\Delta}{=} 0\quad (\text{for all}\ J),
\end{align}
as considered in the literature. One then finds that all the above symmetries are gauge redundancies.
Indeed, it is easy to evaluate charges corresponding to these boundary symmetries as
\begin{align}\label{eq:Q_0}
    \delta Q_0[\alpha] 
    &= I_\alpha\Omega =\frac{1}{2\pi}\delta\ab[\int_{\del\Sigma}\alpha
    G-p\,d\alpha\wedge a-d\alpha\wedge\sum_l v_la^l] \notag\\
    &= \frac{1}{2\pi}\delta\ab(\int_{\del\Sigma}\alpha B_0),
\end{align}
and 
\begin{align}\label{eq:Q_j}
    \delta Q_j[\alpha^j]
    &= I_{\alpha^j}\Omega = -\frac{1}{2\pi}\delta\ab[\int_{\del\Sigma}d\alpha^j \wedge \ab(v_j a + \sum_{l}k_{jl}a^l)] \notag\\
    &= \frac{1}{2\pi}\delta\ab(\int_{\del\Sigma}\alpha^jB_j).
\end{align}
These variations vanish because the corresponding charges are fixed by the boundary conditions.
Thus, we conclude that the choice \eqref{allB=0} makes these symmetry transformations gauge redundant.
The Dirichlet boundary conditions $\delta da=0,\ \delta da^j=0$ likewise do not give rise to physical boundary symmetries.

For simpler boundary actions, boundary conditions have recently been identified under which some of the shift symmetries of edge modes become physical \cite{Ball:2024hqe, Araujo-Regado:2024dpr, Shimizu:2026zug}.
The theory considered in \cite{Ball:2024hqe,Araujo-Regado:2024dpr} (referred to as the Neumann case in \cite{Shimizu:2026zug}) is equivalent to the $K=0$ case without the theta term.
Also, the Dirichlet case in \cite{Shimizu:2026zug} is equivalent to the case of
\begin{align}
    K =
    \begin{pmatrix}
        0 & 1 \\
        1 & 0
    \end{pmatrix},
\end{align}
without the theta term, although a background boundary gauge field is also introduced in \cite{Shimizu:2026zug}.\footnote{Inclusion of the background boundary gauge field is straightforward, and we omit it in the present paper.}
For the general action \eqref{eq:general action}, there are many more choices of boundary conditions than those considered in the works \cite{Ball:2024hqe, Araujo-Regado:2024dpr, Shimizu:2026zug}. 
In other words, we can consider the $\mathrm{SL}(2,\mathbb Z)$ generalizations of the DEM (or soft) boundary conditions in \cite{Ball:2024hqe, Araujo-Regado:2024dpr, Shimizu:2026zug}.  
We can classify these choices according to which charges are made physical.
According to \cite{Ball:2024hqe,Araujo-Regado:2024dpr}, we can obtain a physical charge by fixing the temporal component of a (dressed) boundary gauge field and the temporal-angular components of the variable conjugate to the boundary gauge field.\footnote{
Here we describe only the so-called dynamical edge mode condition, also known as the soft Neumann boundary condition.
Other types include soft Dirichlet and soft Robin boundary conditions, but we will not investigate them in detail. They share with the boundary condition considered here the feature that temporal and spatial components are treated separately.
}

For instance, let us consider a boundary condition
\begin{align}\label{eq:generalized DEM}
    \delta a_t^J \overset{\Delta}{=} 0,\quad (B_J)_{bt}\overset{\Delta}{=}0 \quad (\text{for all}\,J).
\end{align}
Under this boundary condition, the variations of the charges associated with the local shift transformations $\phi^J \to \phi^J +\alpha^J(\Omega)$\footnote{The parameters $\alpha^J$ must be independent of $t$ so that the transformations preserve the boundary conditions $\delta a_t^J \overset{\Delta}{=} 0$.} are still given by \eqref{eq:Q_0} and \eqref{eq:Q_j}:
\begin{align}
    \delta Q_J[\alpha^J] 
    = \frac{1}{2\pi}\delta\ab(\int_{\del\Sigma}\alpha^J B_J).
\end{align}
Note that they can be written in components as
\begin{align}
    \delta Q_J[\alpha^J] = \frac{1}{2\pi}\delta\ab(\int_{\del\Sigma}\alpha^J \frac{1}{2}(B_J)_{ab}\,d\Omega^a\wedge d\Omega^b)
\end{align}
by taking $\del\Sigma$ to be a time slice.
Since the boundary condition \eqref{eq:generalized DEM} does not fix $(B_J)_{ab}$, these charges need not vanish.
In other words, the boundary condition \eqref{eq:generalized DEM} makes the transformation $\phi^J\to\phi^J+\alpha^J(\Omega)$ and the charges 
\begin{equation}
Q_J[\alpha^J]=  \frac{1}{2\pi}\int_{\del\Sigma}\alpha^J B_J
\end{equation}
physical.
These charges are conserved because the associated currents, $*\mathfrak{J}_J[\alpha^J] = \frac{1}{2\pi}\alpha^J B_J$, are conserved, as shown below:
\begin{align}
    d*\mathfrak{J}_J[\alpha^J]
    &= \frac{1}{2\pi}d(\alpha^J(\Omega)B_J) 
    = \frac{1}{2\pi}\ab[d \alpha^J\wedge B_J +\alpha^J dB_J] 
    = 0.
\end{align}
Here, $d \alpha^J\wedge B_J=0$ because both forms are spatial on the two-dimensional spatial slices of $\Delta$, with the spatial character of $B_J$ following from the boundary condition \eqref{eq:generalized DEM}. We have also used $dB_J=0$, which follows from the bulk equation of motion $d G=0$.
Thus, these charges are topological on $\Delta$ as discussed in \cite{Shimizu:2026zug}.
It is worth noting that they can be decomposed as
\begin{align}
    Q_0[\alpha^0] &= Q_e[\alpha^0] + \ab(p-\frac{\theta}{2\pi})Q_m[\alpha^0] + \sum_l v_lQ_{\mathrm{CS}}^l[\alpha^0], \label{eq:charge Q_0} \\
    Q_j[\alpha^j] &= v_j Q_m[\alpha^j] + \sum_{l}k_{jl}Q_{\mathrm{CS}}^l[\alpha^j] \label{eq:charge Q_j}
\end{align}
where we have defined\footnote{Note that the standard quantized electric charge is $Q_e[\alpha^0] -\frac{\theta}{2\pi}Q_m[\alpha^0]$ in the presence of the $\theta$ term because the ``electric displacement field'' satisfying the Gauss law on $\Delta$ is $* F - \frac{e^2 \theta}{4\pi^2} F$, when we ignore the boundary couplings.}
\begin{align}
    Q_e[\alpha] = \frac{1}{e^2}\int_{\del\Sigma}\alpha*F,\quad Q_m[\alpha] = \frac{1}{2\pi}\int_{\del\Sigma}\alpha F, \quad Q_{\mathrm{CS}}^l[\alpha] = \frac{1}{2\pi}\int_{\del\Sigma}\alpha da^l.
\end{align}
The ``charges'' $Q_{e,m,\mathrm{CS}}$ are not conserved and are not topological operators, although their linear combinations $Q_J$ are conserved and topological on the boundary $\Delta$.

The Poisson brackets of the charges $Q_J$ are
\begin{align}
\label{eq:comK}
    \{Q_I[\alpha], Q_J[\beta]\}=\frac{K_{IJ}}{2\pi}\int_{\partial\Sigma} d\alpha \wedge d\beta.
\end{align}
For non-singular parameters $\alpha, \beta$, these central charges vanish:
\begin{align}
    \{Q_I[\alpha], Q_J[\beta]\}=0.
\end{align}
Although the charges $Q_I$ are physical and it is therefore meaningful to discuss their commutators, the central charges vanish for non-singular parameters.
If we allow singular parameters $\alpha$ and $\beta$ as discussed in \cite{Freidel:2018fsk, Geiller:2021gdk} (and \cite{Hosseinzadeh:2018dkh} throughout Minkowski spacetime), we may obtain non-vanishing central charges.
However, singular transformations can change the action, as argued in \cite{Mathieu:2020fwg, Shimizu:2026zug}, so $Q_J[\alpha]$ for singular $\alpha$ are not symmetry charges. It therefore remains unclear whether the corresponding central charges are physically meaningful.

Electric and magnetic edge modes in pure Maxwell theory are boundary degrees of freedom associated with large gauge transformations of the gauge field and the dual gauge field that act nontrivially at the boundary.
They therefore correspond to the pure-gauge components of the gauge field and its dual, as discussed in \cite{Shimizu:2026zug}, and can be interpreted as endpoints of a Wilson line and a 't Hooft line of the bulk gauge field. 
In a generic system \eqref{eq:general action}, the edge mode $\phi$ can be interpreted as the endpoint of a Wilson line of the bulk gauge field, and the expression for charge $Q_0$ \eqref{eq:charge Q_0} shows that $\phi$ is generically charged by $Q_{e, m, \mathrm{CS}}$. 
In other words, $Q_{e, m, \mathrm{CS}}$ can act on Wilson lines of $A$ attached to the boundary.
In this sense, $\phi$ can be regarded as an endpoint of a dyonic line.
Similarly, $\phi_j$ can be interpreted as an endpoint of Wilson lines of $A_j$ and in general they are magnetically charged with respect to the bulk gauge field due to the term $Q_m$ in \eqref{eq:charge Q_j}.
In addition, some $\phi_j$ may correspond to the dual counterpart of $\phi_0$ under the $\mathrm{SL}(2,\mathbb Z)$ transformations as we will see in \cref{sec:transf laws}. Since the Wilson lines are mapped to the dyonic lines under  $\mathrm{SL}(2,\mathbb Z)$ duality, some $\phi_j$ can be interpreted as endpoints of the dyonic lines.

The variational principle \eqref{eq:requirement} allows us to take another boundary condition. 
It is possible to turn off some of the charges by imposing mixed boundary conditions as
\begin{gather}
    \delta a^{s}_t = 0,\quad (B_s)_{bt}=0\quad (s=\sigma_0,\cdots,\sigma_{n'}), \label{eq:mixed boundary condition1}\\
    B_s =0, \quad (s=\sigma_{n'+1},\cdots,\sigma_n),\label{eq:mixed boundary condition2}
\end{gather}
where $(\sigma_0,\ldots,\sigma_n)$ is a permutation of $(0,\ldots,n)$.
These boundary conditions make the shifts $\phi^s\to\phi^s+\alpha^s(\Omega)\,(s=\sigma_0,\cdots,\sigma_{n'})$ physical, while rendering the remaining shifts $\phi^s\to\phi^s+\alpha^s(t, \Omega)\,(s=\sigma_{n'+1},\cdots,\sigma_n)$ gauge redundancies.
The physical charges $Q_J$, and hence the corresponding symmetries, are mutually independent.
Therefore, the number of physical charges
is equal to that of \eqref{eq:generalized DEM}-type boundary conditions we impose, which means that the system with the boundary condition \eqref{eq:mixed boundary condition1} and \eqref{eq:mixed boundary condition2} has $n'+1$ physical charges.
Also, we can say that the system has the same number ($n'+1$ in this case) of edge modes.

Let us consider the specific case with $\theta=0$ and
\begin{align}\label{eq:dyonic-K}
    K =
    \begin{pmatrix}
        0 & 1 \\
        1 & 0
    \end{pmatrix}.
\end{align}
This theory is equivalent to the system considered in Section 4.2 of \cite{Shimizu:2026zug}, where the authors impose the following boundary conditions:
\begin{gather}
    \delta a^0 \overset{\Delta}{=} \delta\Lambda_a d\Omega^a + d\delta\lambda,\quad (B_0)_{bt} \overset{\Delta}{=} 0, \\
    \delta a^1_t \overset{\Delta}{=} 0,\quad (B_1)_{bt} \overset{\Delta}{=} 0.
\end{gather}
This choice differs slightly from \eqref{eq:generalized DEM} because the freedom in $\delta \lambda$ allows $\delta a^0_t \neq 0$.
It satisfies \eqref{eq:requirement} with nonzero $dc$ in \eqref{eq:dc}.
In this case, the charge $Q_0$ associated with the shift symmetry $\phi^0 \to \phi^0+\alpha^0$ vanishes and thus the edge mode $\phi^0$ is gauge redundant \cite{Shimizu:2026zug}.

However, the situation changes if we impose the boundary condition \eqref{eq:generalized DEM}, i.e.,
\begin{align}\label{eq:dyonic-bc}
    \delta a_t^0 \overset{\Delta}{=} 0,\quad (B_0)_{bt}\overset{\Delta}{=}0,\quad \delta a_t^1 \overset{\Delta}{=} 0,\quad (B_1)_{bt}\overset{\Delta}{=}0.
\end{align}
We then have physical electric and magnetic charges $Q_0=Q_e+Q_\mathrm{CS}^1$ and $Q_1=Q_m$.
We are therefore justified in considering the commutation relation between these charges:
\begin{align}
    \{Q_e[\alpha]+Q_{\mathrm{CS}}[\alpha],Q_m[\hat\alpha]\} = \frac{1}{2\pi}\int_{\del\Sigma} d\alpha\wedge d\hat\alpha.
\end{align}
This is a specific example of \eqref{eq:comK}.
As discussed there, this commutator vanishes for non-singular parameters.
Thus, for non-singular parameters, the central charge in the algebra of these physical charges vanishes.
Note that this commutation relation differs slightly from that considered in \cite{Freidel:2018fsk, Geiller:2021gdk} due to the additional Chern--Simons contributions $Q_{\mathrm{CS}}$ in the bracket.
In this construction, making the magnetic charge physical inevitably introduces a Chern--Simons contribution to $Q_0$, as can be seen from \eqref{eq:charge Q_0} and \eqref{eq:charge Q_j}.

\section{Transformation laws of boundary conditions and charges}\label{sec:transf laws}
Let us consider how the boundary condition \eqref{eq:generalized DEM} and the associated physical charges $Q_J$ transform under the $S$ and $T$ transformations.

\subsection{\texorpdfstring{$S$}{S}  transformation}\label{subsec:Strsf}
\begin{figure}
    \centering
    \includegraphics[
        width=0.7\linewidth,
        trim=550bp 510bp 700bp 215bp,
        clip
    ]{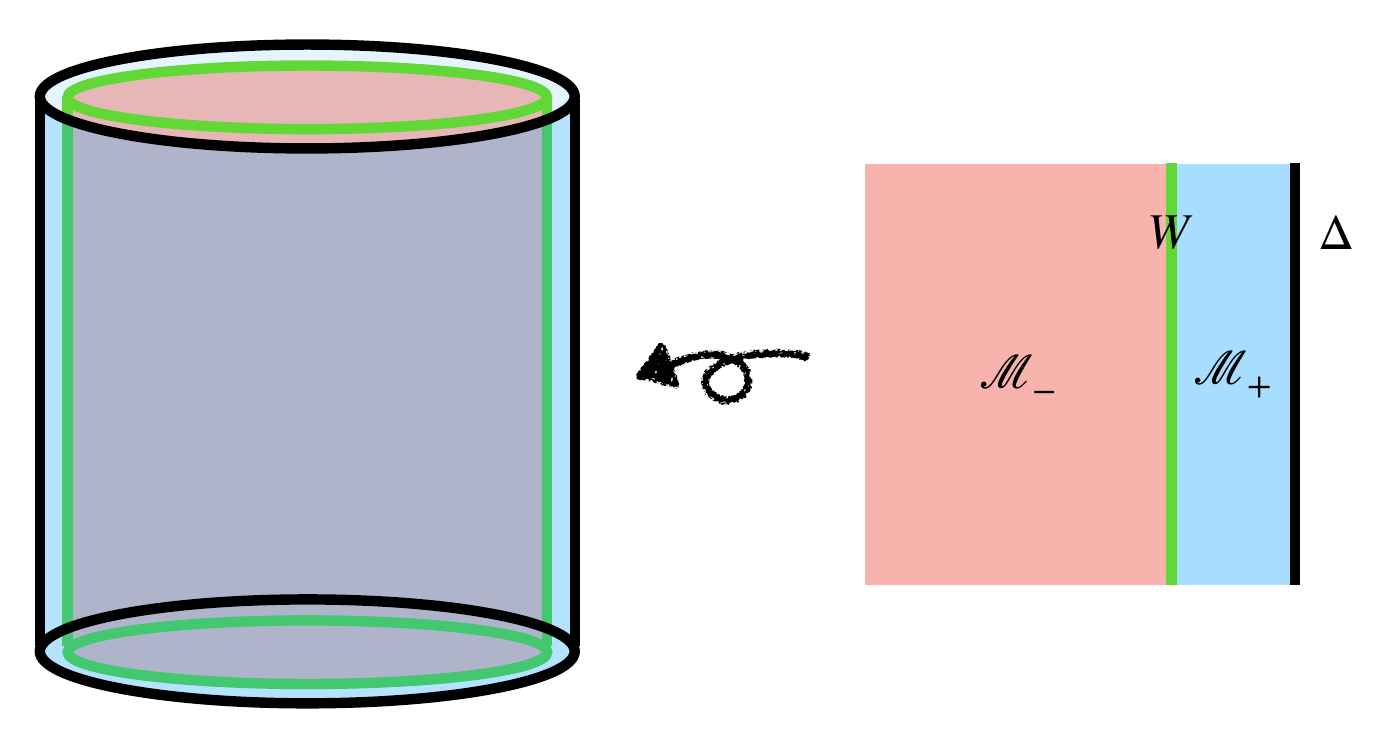}
    \caption{A spacetime with an $S$ wall $W$ and a boundary $\Delta$. The regions inside and outside the wall are denoted by $\mathcal{M}_-$ and $\mathcal{M}_+$, respectively.}
    \label{fig:spacetime with wall}
\end{figure}

We first consider the $S$ transformation by fusing a dressed $S$ wall with $\Delta$.
The original action is given by \eqref{eq:general action}.
Let us place the wall at $W=\{r=R-\epsilon\}$ with $\epsilon>0$, separating the interior $\mathcal{M}_-=\{r<R-\epsilon\}$ from the neighborhood $\mathcal{M}_+=\{R-\epsilon<r<R\}$ of $\Delta$, as shown in \cref{fig:spacetime with wall}.
Let $A_\pm$ and $(e_\pm,\theta_\pm)$ be the bulk gauge fields and couplings in $\mathcal{M}_\pm$, respectively, and define
\begin{align}\label{eq:duality-G}
    F_\pm=dA_\pm,\qquad
    G_\pm=\frac{2\pi}{e_\pm^2}*F_\pm-\frac{\theta_\pm}{2\pi}F_\pm.
\end{align}
The dressed fields on the wall and the original boundary are
\begin{align}
    a_\pm^W=A_\pm|_{W}+d\phi_\pm^W,\qquad
    a_+^\Delta=A_+|_\Delta+d\phi_+^\Delta,
\end{align}
where $\phi_\pm^W$ and $\phi_+^\Delta$ are edge modes on the indicated surfaces.
On $\Delta$, we also have the original boundary modes $(A^j, \phi^j)$ ($j=1, \cdots, n$).
We choose the orientation of $W$ to agree with that of $\Delta$, so that the $S$ wall action is
\begin{align}
    S_W=\frac{1}{2\pi}\int_{W}a_-^W\wedge da_+^W.
\end{align}
The contribution localized on $W$ to the variation of the total action is
\begin{align}
    \delta S|_{W}
    =\frac{1}{2\pi}\int_{W}\left[
    \delta A_-\wedge\ab(G_-+F_+)
    +\delta A_+\wedge\ab(-G_++F_-)\right].
\end{align}
Variations of the edge modes give no additional contributions because $dF_\pm=0$.
Thus the wall matching conditions are
\begin{align}\label{eq:S-wall-matching}
    G_-+F_+\overset{W}{=}0,\qquad
    G_+-F_-\overset{W}{=}0.
\end{align}

Fusing the wall with $\Delta$ by taking the limit $\epsilon\to0$, we obtain the following new dressed boundary fields:
\begin{align}
    \widetilde a^0=A_-|_\Delta+d\phi_-^W,\qquad
    \widetilde a^1=a_+^\Delta,\qquad
    \widetilde a^{j+1}=a^j\quad (j=1,\ldots,n).
\end{align}
Thus $\widetilde a^1$ corresponds to the original dressed gauge field $a_+$.
The wall term becomes $\frac{1}{2\pi}\int_\Delta\widetilde a^0\wedge d\widetilde a^1$, giving the matrix in \eqref{eq:duality-K},
\begin{align}
    \widetilde{K}=\begin{pmatrix}
        0&1&0\\
        1&p&\bm{v}^{\mathsf{T}}\\
        0&\bm{v}&k
    \end{pmatrix}.
\end{align}

The corresponding new boundary two-forms are
\begin{align}
    \widetilde B_0&=G_-+d\widetilde a^1,\qquad 
    \widetilde B_1=d\widetilde a^0+p\,d\widetilde a^1+\sum_{j=1}^n v_jd\widetilde a^{j+1},\nn
    \widetilde B_{j+1}&=v_jd\widetilde a^1+\sum_{l=1}^n k_{jl}d\widetilde a^{l+1}\quad (j=1,\ldots,n).
\end{align}
Let $B_J^{(+)}$ denote $B_J$ in the original theory, with $a=a_+^\Delta$.
Using the matching condition \eqref{eq:S-wall-matching} and $d\widetilde a^0=F_-|_\Delta$, we obtain the following relations between the boundary two-forms before and after the $S$ transformation:
\begin{align}
    \widetilde B_0=0,\qquad
    B_0^{(+)}=\widetilde B_1,\qquad
    B_j^{(+)}=\widetilde B_{j+1}\quad (j=1,\ldots,n).
\end{align}

Hence the original boundary condition \eqref{eq:generalized DEM} leads to the following boundary condition for the new fields:
\begin{align}\label{eq:S-image-DEM}
    \widetilde B_0\overset{\Delta}{=}0,\qquad
    \delta\widetilde a_t^i\overset{\Delta}{=}0,\qquad
    (\widetilde B_i)_{bt}\overset{\Delta}{=}0
    \quad (i=1,\ldots,n+1).
\end{align}
In particular, the conditions involving the new bulk field $A_-$ are
\begin{align}
    \frac{2\pi}{e_-^2}*F_--\frac{\theta_-}{2\pi}F_-+d\widetilde a^1&\overset{\Delta}{=}0,\notag\\
    \ab(F_-+p\,d\widetilde a^1+\sum_{j=1}^n v_jd\widetilde a^{j+1})_{bt}&\overset{\Delta}{=}0,
\end{align}
while the condition $\delta\widetilde a_t^0=0$ is not imposed.
The resulting boundary condition belongs to the mixed class defined by \eqref{eq:mixed boundary condition1} and \eqref{eq:mixed boundary condition2}.

Let $\widetilde{\phi}^J$ denote the new edge modes. They are related to the original edge modes by $\widetilde\phi^0=\phi_-^W$, $\widetilde\phi^1=\phi_+$, and $\widetilde\phi^{j+1}=\phi^j$.
As mentioned below \eqref{eq:mixed boundary condition2}, some of the edge modes are physical.
In the case of \eqref{eq:S-image-DEM},
the modes $\widetilde{\phi}^i$ $(i=1,\ldots,n+1)$ are physical, and the corresponding shifts $\widetilde{\phi}^i \to \widetilde{\phi}^i+\alpha^i(\Omega)$ are physical symmetries.
The associated charges are
\begin{align}
    \widetilde Q_i[\alpha^i]
    =\frac{1}{2\pi}\ab(\int_{\del\Sigma}\alpha^i\widetilde B_i).
\end{align}
On the other hand, the shift of $\widetilde\phi^0$ is redundant because $\delta\widetilde Q_0[\alpha^0]=\frac{1}{2\pi}\int_{\del\Sigma}\alpha^0\delta\widetilde B_0=0$.
Thus, after the $S$ transformation, we obtain an additional edge mode but it is a redundant degree of freedom. Hence the numbers of conserved charges and physical edge modes are unchanged under the $S$ transformation.

Using $d\widetilde a^1=-G_-$, the physical charges can be written as
\begin{align}\label{eq:S-physical-charges}
    \widetilde Q_1[\alpha]
    &=\frac{1}{2\pi}\int_{\del\Sigma}\alpha\ab(F_--pG_-+\sum_{j=1}^n v_jd\widetilde a^{j+1}),\notag\\
    \widetilde Q_{j+1}[\alpha]
    &=\frac{1}{2\pi}\int_{\del\Sigma}\alpha\ab(-v_jG_-+\sum_{l=1}^n k_{jl}d\widetilde a^{l+1})
    \quad (j=1,\ldots,n),
\end{align}
where $\alpha=\alpha(\Omega)$ is independent of time $t$.
These charges are topological on $\Delta$ as discussed in the previous section.
Let $Q_J^{(+)}[\alpha]=\frac{1}{2\pi}\int_{\del\Sigma}\alpha B_J^{(+)}$ denote the original charges before fusion.
The wall matching conditions $F_-=G_+$ and $G_-=-F_+$, together with $\widetilde a^{j+1}=a^j$, lead to the following $S$ transformation law for the charges:
\begin{align}\label{eq:S-charge-map}
    \widetilde Q_{J+1}[\alpha]=Q_J^{(+)}[\alpha] \quad  (J=0,\ldots,n).
\end{align}

Since $S^2=-1$, we expect to recover the charge-conjugated boundary couplings and boundary conditions after two successive $S$ transformations. We confirm this in Appendix~\ref{app:S-wall}.

\subsection{\texorpdfstring{$T$}{T} transformation}

We next consider the $T$ transformation. Unlike the $S$ transformation, the $T$ transformation does not introduce an additional boundary gauge field. 
We denote quantities after the $T$ transformation by tildes. 
Under the $T$ transformation, the bulk coupling changes as 
\begin{align}
    \tau \longrightarrow \widetilde{\tau}=\tau+1,
\end{align}
that is, $\widetilde{\theta}=\theta+2\pi$, $\widetilde{e}=e$,
while the gauge field remains unchanged: $\widetilde{A}=A$.
Therefore, we have 
\begin{align}\label{Ttransformation_FG}
    \widetilde{F}=F, \qquad \widetilde{G}=
    \frac{2\pi}{e^2}\ast F
    -\frac{\widetilde{\theta}}{2\pi}F=G-F.
\end{align}

At the same time, the boundary matrix transforms as
\begin{align}
    K=
    \begin{pmatrix}
        p & v^{T} \\
        v & k
    \end{pmatrix}
    \longrightarrow
    \widetilde{K}=
    \begin{pmatrix}
        p+1 & v^{T} \\
        v & k
    \end{pmatrix}.
\end{align}
Since no new boundary gauge field is introduced, we identify
\begin{align}
    \widetilde{a}_J=a_J,
    \qquad
    (J=0,\ldots,n).
\end{align}

The boundary two-forms after the $T$ transformation are
\begin{align}
    \widetilde{B}_0
    &=
    \widetilde{G}
    +(p+1)da
    +\sum_{l=1}^{n}v_l\,da_l
    =G-F
    +(p+1)da
    +\sum_{l=1}^{n}v_l\,da_l
    \notag\\
    &=
    B_0,
\end{align}
where we used $da=F|_{\Delta}$, and
\begin{align}
    \widetilde{B}_j
    =B_j,
    \qquad
    (j=1,\ldots, n).
\end{align}
Thus, the boundary two-forms are invariant under the $T$ transformation, 
\begin{align}
    \widetilde{B}_J=B_J,
    \qquad
    (J=0,\ldots,n).
\end{align}
The boundary condition \eqref{eq:generalized DEM} is therefore also invariant:
\begin{align}
    \left.\delta\widetilde{a}_{Jt}\right|_{\Delta}=0,
    \qquad
    \left.(\widetilde{B}_J)_{bt}\right|_{\Delta}=0,
    \qquad
    (J=0,\ldots,n).
\end{align}

The physical charges after the $T$ transformation are
\begin{align}
    \widetilde{Q}_J[\alpha]
    =
    \frac{1}{2\pi}
    \int_{\partial\Sigma}
    \alpha\,\widetilde{B}_J.
\end{align}
Using $\widetilde{B}_J=B_J$, we obtain the $T$ transformation law for the conserved charges:
\begin{align}
    \widetilde{Q}_J[\alpha]
    =
    Q_J[\alpha],
    \qquad
    (J=0,\ldots,n).
\end{align}
Hence the physical charges are invariant under the $T$ transformation.
This invariance refers to the conserved boundary charges $Q_J$.
By contrast, the electric displacement flux defined using only the bulk fields transforms nontrivially under $T$ as \eqref{Ttransformation_FG}.
The naive electric charge obtained by ignoring the boundary action is
\begin{align}
   Q_e^{\text{bulk}}= \frac{1}{2\pi}\int G = \frac{1}{2\pi}\int \left(\frac{2\pi}{e^2}*F - \frac{\theta}{2\pi}F\right),
\end{align}
which changes under $\theta \to \theta + 2\pi$ as
\begin{align}
    Q_e^{\text{bulk}} \longrightarrow Q_e^{\text{bulk}} -  \frac{1}{2\pi}\int F =  Q_e^{\text{bulk}} -Q_m.
\end{align}
This charge mixing $Q_e^{\text{bulk}} \to Q_e^{\text{bulk}} - Q_m$ reflects the Witten effect \cite{Witten:1979ey} (where the sign depends on the sign convention for the $\theta$ term).
However, the conserved charge $Q_0$ receives additional contributions from the boundary action, as shown in \eqref{eq:charge Q_0}. The parameter $\theta$ appears in $Q_0$ only through the combination $p-\theta/(2\pi)$, which is invariant under the simultaneous shifts $\theta \to \theta + 2\pi$, $p \to p+1$.

\section{Summary and discussion}
In this paper, we considered a general Abelian gauge theory in a region with a boundary at finite radius, where the bulk Maxwell theory with a theta term is coupled to a general Chern--Simons theory on the boundary.
This setup provides a convenient framework for studying how the $\mathrm{SL}(2,\mathbb Z)$ transformations act on boundary conditions and boundary symmetries.
We studied possible boundary conditions for the general action \eqref{eq:general action}.
Although the theory appears to admit boundary symmetries, standard boundary conditions, such as the perfect magnetic conductor (PMC) and perfect electric conductor (PEC) boundary conditions, do not, does not allow any physical boundary symmetries.
We considered modified boundary conditions as in \cite{Ball:2024hqe, Araujo-Regado:2024dpr, Shimizu:2026zug} which allow many physical boundary symmetries and genuine physical charges.
Since the general action \eqref{eq:general action} can be obtained by $\mathrm{SL}(2,\mathbb Z)$ transformations from the standard action without the boundary couplings, the new boundary conditions can be regarded as $\mathrm{SL}(2,\mathbb Z)$ generalizations of the dynamical edge mode boundary condition or soft boundary conditions in \cite{Ball:2024hqe, Araujo-Regado:2024dpr, Shimizu:2026zug}.
The resulting charges $Q_J$ can be expressed as linear combinations of $Q_{e, m, \mathrm{CS}}$ as in \eqref{eq:charge Q_0} and \eqref{eq:charge Q_j}.
This means that naive electric and magnetic charges $Q_{e, m, \mathrm{CS}}$ are not individually conserved for generic boundary couplings, whereas certain linear combinations $Q_J$ are conserved.
Each physical symmetry generated by $Q_J$ has an associated edge mode.
In addition, we discovered a new boundary condition under which both electric and magnetic charges can be physical.
This enables us to discuss the central charge \cite{Hosseinzadeh:2018dkh,Freidel:2018fsk,Geiller:2021gdk} of the commutation relation between electric and magnetic charges which was previously obscure since there is no known boundary condition where both charges are physical simultaneously.
Nevertheless, it remains slightly unclear whether this central extension has a physically meaningful interpretation, since the central charge takes the form of an integral of a total derivative over a closed surface.
This integral can be nonzero only for singular transformations, which typically introduce line operators on the boundary.
We leave a detailed investigation of this issue for future work.
In \cref{sec:transf laws}, we studied the $S$ and $T$ transformations of our boundary conditions and the associated physical charges.

Finally, let us comment on the case with charged matter.
For simplicity, we have not included charged matter in our analysis. However, it can be incorporated as long as it does not reach the boundary and its current couples to the gauge field through $A\wedge*j$.
A specific case with charged matter was already investigated in \cite{Shimizu:2026zug}.

\acknowledgments
We thank an anonymous referee of \cite{Shimizu:2026zug} for asking about $\mathrm{SL}(2,\mathbb Z)$ duality.
KS was supported by Grant-in-Aid for JSPS Fellows No.26KJ1405.
SS acknowledges support from JSPS KAKENHI Grant Numbers JP22H05115, 26K17145 and JST BOOST Program Japan Grant Number JPMJBY24E0.

\appendix
\section{Two successive \texorpdfstring{$S$}{S} transformations}\label{app:S-wall}

The $S$ transformation maps a theory with a boundary condition into one with another boundary condition.
We expect two successive $S$ transformations to recover the original theory up to charge conjugation even in the presence of physical edge modes, since $S^2=-1$ in $\mathrm{SL}(2,\mathbb Z)$.\footnote{More precisely, we have $S^2:\, A \rightarrow -A$, which is charge conjugation, denoted by $C$. Thus, after two successive $S$ transformations, we obtain the charge-conjugate theory.}
We confirm this fact for a specific case because the extension to the generic case is straightforward.

Let us consider the case $\theta=0$ and
\begin{align}
    K =
    \begin{pmatrix}
        0 & 1 \\
        1 & 0
    \end{pmatrix}
\end{align}
with the boundary condition \eqref{eq:dyonic-bc}.
We label quantities after $m$ transformations by $(m)$, with $m=0,1,2$.
The bulk fields $A_{(m)}$, $F_{(m)}$, and $G_{(m)}$ are defined as in \eqref{eq:duality-G}, with the couplings of the corresponding theory.
The dressed boundary fields are 
\begin{align}
    a_{(m)}^0=A_{(m)}|_\Delta+d\phi_{(m)}^0,
\end{align}
where $\phi_{(m)}^0$ is the edge mode associated with the bulk field $A_{(m)}$.
We also have dressed gauge fields
\begin{align}
    a^i_{(m)} \qquad (i=1,\cdots, m+1).
\end{align}
The boundary two-form \eqref{eq:B-matrix} is given by
\begin{align}
    B_J^{(m)}=\delta_{J0}\,G_{(m)}
    +\sum_{L=0}^{m+1} K^{(m)}_{JL}da_{(m)}^L.
\end{align}

The matching condition \eqref{eq:S-wall-matching} implies that the bulk field strengths transform as follows:
\begin{align}
    (F_{(1)},G_{(1)})=(G_{(0)},-F_{(0)}),\qquad
    (F_{(2)},G_{(2)})=(-F_{(0)},-G_{(0)}).
\end{align}
The second equation shows that the bulk fields undergo charge conjugation $C$, while the coupling remains unchanged: $\tau_{(0)}=\tau_{(2)}$.
We will see that the boundary gauge fields and boundary charges also undergo charge conjugation after two successive $S$ transformations.
Initially, we have
\begin{align}
    B_0^{(0)}=G_{(0)}+da_{(0)}^1,\qquad
    B_1^{(0)}=da_{(0)}^0,
\end{align}
with the boundary condition
\begin{align}
    \delta a_{(0)t}^J=0,\quad (B_J^{(0)})_{bt}=0\quad (J=0,1).
\end{align}

After one $S$ transformation, the matrix $K$ and boundary two-forms are given by
\begin{align}
    K^{(1)}=\begin{pmatrix}
        0&1&0\\
        1&0&1\\
        0&1&0
    \end{pmatrix},\qquad
    \begin{pmatrix}B_0^{(1)}\\B_1^{(1)}\\B_2^{(1)}\end{pmatrix}
    =\begin{pmatrix}
        G_{(1)}+da_{(1)}^1\\
        d(a_{(1)}^0+a_{(1)}^2)\\
        da_{(1)}^1
    \end{pmatrix}.
\end{align}
As computed in \cref{subsec:Strsf}, we have 
\begin{align}
    a_{(1)}^{i+1}=a_{(0)}^i,\quad B_0^{(1)}=0, \quad B_{i+1}^{(1)}=B_i^{(0)} \quad (i=0,1).
\end{align}
The transformed boundary condition is
\begin{align}\label{eq:dyonic-S-bc}
    B_0^{(1)}=0,\qquad
    \delta a_{(1)t}^i=0,\qquad (B_i^{(1)})_{bt}=0\quad (i=1,2).
\end{align}
Applying $S$ once more gives
\begin{align}
    K^{(2)}=\begin{pmatrix}
        0&1&0&0\\
        1&0&1&0\\
        0&1&0&1\\
        0&0&1&0
    \end{pmatrix},\qquad
    \begin{pmatrix}B_0^{(2)}\\B_1^{(2)}\\B_2^{(2)}\\B_3^{(2)}\end{pmatrix}
    =\begin{pmatrix}
        G_{(2)}+da_{(2)}^1\\
        d(a_{(2)}^0+a_{(2)}^2)\\
        d(a_{(2)}^1+a_{(2)}^3)\\
        da_{(2)}^2
    \end{pmatrix}.
\end{align}
Since $B_0^{(2)}=0$ and $B_{i+1}^{(2)}=B_i^{(1)}$, the boundary condition \eqref{eq:dyonic-S-bc} leads to
\begin{align}\label{eq:dyonic-S2-bc}
    B_0^{(2)}=B_1^{(2)}=0,\qquad
    \delta a_{(2)t}^i=0,\qquad (B_i^{(2)})_{bt}=0\quad (i=2,3).
\end{align}

We can reduce the boundary theory using \eqref{eq:dyonic-S2-bc}.
The condition $0=B_1^{(2)}=d(a_{(2)}^0+a_{(2)}^2)$ implies
\begin{align}
    a_{(2)}^0+a_{(2)}^2=d\sigma.
\end{align}
Here, $\sigma$ is a scalar on $\Delta$; such a scalar exists because every closed one-form on $\Delta\simeq\mathbb R\times S^2$ is exact.
Since $B_0^{(2)}=0$, the shift of $a_{(2)}^0$ is redundant.
Using $a_{(2)}^0\to a_{(2)}^0-d\sigma$, we can choose $a_{(2)}^2=-a_{(2)}^0$ and define
\begin{align}
    a_C^0\equiv a_{(2)}^0=-a_{(0)}^0,\qquad
    a_C^1\equiv -a_{(2)}^3=-a_{(0)}^1.
\end{align}
The boundary action specified by $K^{(2)}$ then becomes
\begin{align}
    S_{\Delta,C}=\frac{1}{2\pi}\int_\Delta a_C^0\wedge da_C^1,\qquad
    K_C=\begin{pmatrix}0&1\\1&0\end{pmatrix}=K.
\end{align}
This is the original boundary coupling with its bulk field charge conjugated.
The reduced boundary two-forms satisfy
\begin{align}
    B_0^C&=G_{(2)}^C+da_C^1=-G_{(0)}-da_{(0)}^1=-B_0^{(0)},\notag\\
    B_1^C&=da_C^0=-da_{(0)}^0=-B_1^{(0)}.
\end{align}
Thus, the reduced theory is related to the original theory by charge conjugation, with $a^J_C=-a^J_{(0)}$ and $B_J^C=-B_J^{(0)}$.
The boundary condition \eqref{eq:dyonic-S2-bc} leads to
\begin{align}
    \delta a_{Ct}^J=0,\qquad (B_J^C)_{bt}=0\quad (J=0,1),
\end{align}
which is the same as \eqref{eq:dyonic-bc} because it is invariant under charge conjugation.

In the theory obtained after two $S$ transformations, the physical non-vanishing charges are associated with the shifts of $a_{(2)}^{2,3}$, while the shifts of $a_{(2)}^{0, 1}$ are gauge redundant.
However, in the reduced theory, to preserve $a_{(2)}^2=-a_{(2)}^0$, a physical shift $a_{(2)}^2\to a_{(2)}^2+d\alpha$ is accompanied by the redundant shift $a_{(2)}^0\to a_{(2)}^0-d\alpha$, with $\alpha=\alpha(\Omega)$.
This combined transformation therefore induces a physical shift of $a_C^0$, and similarly the shift of $a_{(2)}^3$ induces that of $a_C^1$.
In the reduced theory, the physical charges are 
$Q_J^C[\alpha]\equiv\frac{1}{2\pi}\int_{\del\Sigma}\alpha B_J^C$, and they are related to the original charges by $Q_0^C[\alpha]=-Q_0[\alpha]$ and $Q_1^C[\alpha]=-Q_1[\alpha]$.
Thus, the two physical charges, the two physical edge modes, and their boundary conditions agree with those of the original theory after charge conjugation.

\bibliography{ref}

@article{Regge:1974zd,
    author = "Regge, Tullio and Teitelboim, Claudio",
    title = "{Role of Surface Integrals in the Hamiltonian Formulation of General Relativity}",
    reportNumber = "Print-74-0988 (IAS,PRINCETON)",
    doi = "10.1016/0003-4916(74)90404-7",
    journal = "Annals Phys.",
    volume = "88",
    pages = "286",
    year = "1974"
}

@article{Benguria:1976in,
    author = "Benguria, R. and Cordero, P. and Teitelboim, C.",
    title = "{Aspects of the Hamiltonian Dynamics of Interacting Gravitational Gauge and Higgs Fields with Applications to Spherical Symmetry}",
    reportNumber = "Print-77-0408 (PRINCETON)",
    doi = "10.1016/0550-3213(77)90426-6",
    journal = "Nucl. Phys. B",
    volume = "122",
    pages = "61--99",
    year = "1977"
}

@article{Brown:1986nw,
    author = "Brown, J. David and Henneaux, M.",
    title = "{Central Charges in the Canonical Realization of Asymptotic Symmetries: An Example from Three-Dimensional Gravity}",
    doi = "10.1007/BF01211590",
    journal = "Commun. Math. Phys.",
    volume = "104",
    pages = "207--226",
    year = "1986"
}

@article{Gervais:1976ec,
    author = "Gervais, J.L. and Sakita, B. and Wadia, S.",
    title = "{The Surface Term in Gauge Theories}",
    doi = "10.1016/0370-2693(76)90467-6",
    journal = "Phys. Lett. B",
    volume = "63",
    pages = "55",
    year = "1976"
}

@article{Wadia:1977qr,
    author = "Wadia, Spenta R.",
    title = "{Hamiltonian Formulation of Nonabelian Gauge Theory with Surface Terms: Applications to the Dyon Solution}",
    reportNumber = "CCNY-HEP 77/4",
    doi = "10.1103/PhysRevD.15.3615",
    journal = "Phys. Rev. D",
    volume = "15",
    pages = "3615",
    year = "1977"
}

@article{Wadia:1976fa,
    author = "Wadia, S. and Yoneya, T.",
    title = "{The Role of Surface Variables in the Vacuum Structure of Yang-Mills Theory}",
    doi = "10.1016/0370-2693(77)90010-7",
    journal = "Phys. Lett. B",
    volume = "66",
    pages = "341--345",
    year = "1977"
}

@article{Balachandran:1991dw,
    author = "Balachandran, A. P. and Bimonte, G. and Gupta, K. S. and Stern, A.",
    title = "{Conformal edge currents in Chern-Simons theories}",
    eprint = "hep-th/9110072",
    archivePrefix = "arXiv",
    reportNumber = "SU-4228-487, INFN-NA-IV-91-12, UAHEP-917, UAHEP-9113",
    doi = "10.1142/S0217751X92002106",
    journal = "Int. J. Mod. Phys. A",
    volume = "7",
    pages = "4655--4670",
    year = "1992"
}

@article{Balachandran:1994up,
    author = "Balachandran, A. P. and Chandar, L. and Momen, Arshad",
    title = "{Edge states in gravity and black hole physics}",
    eprint = "gr-qc/9412019",
    archivePrefix = "arXiv",
    reportNumber = "SU-4240-590",
    doi = "10.1016/0550-3213(95)00622-2",
    journal = "Nucl. Phys. B",
    volume = "461",
    pages = "581--596",
    year = "1996"
}

@article{Kapustin:2009av,
    author = "Kapustin, Anton and Tikhonov, Mikhail",
    title = "{Abelian duality, walls and boundary conditions in diverse dimensions}",
    eprint = "0904.0840",
    archivePrefix = "arXiv",
    primaryClass = "hep-th",
    doi = "10.1088/1126-6708/2009/11/006",
    journal = "JHEP",
    volume = "11",
    pages = "006",
    year = "2009"
}

@inproceedings{Witten:2003ya,
    author = "Witten, Edward",
    title = "{SL(2,Z) action on three-dimensional conformal field theories with Abelian symmetry}",
    booktitle = "{From Fields to Strings: Circumnavigating Theoretical Physics: A Conference in Tribute to Ian Kogan}",
    eprint = "hep-th/0307041",
    archivePrefix = "arXiv",
    pages = "1173--1200",
    month = "7",
    year = "2003"
}

@article{Gaiotto:2008ak,
    author = "Gaiotto, Davide and Witten, Edward",
    title = "{S-Duality of Boundary Conditions In N=4 Super Yang-Mills Theory}",
    eprint = "0807.3720",
    archivePrefix = "arXiv",
    primaryClass = "hep-th",
    doi = "10.4310/ATMP.2009.v13.n3.a5",
    journal = "Adv. Theor. Math. Phys.",
    volume = "13",
    number = "3",
    pages = "721--896",
    year = "2009"
}

@article{Balachandran:2013wsa,
    author = "Balachandran, A. P. and Vaidya, S.",
    title = "{Spontaneous Lorentz Violation in Gauge Theories}",
    eprint = "1302.3406",
    archivePrefix = "arXiv",
    primaryClass = "hep-th",
    doi = "10.1140/epjp/i2013-13118-9",
    journal = "Eur. Phys. J. Plus",
    volume = "128",
    pages = "118",
    year = "2013"
}

@article{Strominger:2013lka,
    author = "Strominger, Andrew",
    title = "{Asymptotic Symmetries of Yang-Mills Theory}",
    eprint = "1308.0589",
    archivePrefix = "arXiv",
    primaryClass = "hep-th",
    doi = "10.1007/JHEP07(2014)151",
    journal = "JHEP",
    volume = "07",
    pages = "151",
    year = "2014"
}

@article{Strominger:2013jfa,
    author = "Strominger, Andrew",
    title = "{On BMS Invariance of Gravitational Scattering}",
    eprint = "1312.2229",
    archivePrefix = "arXiv",
    primaryClass = "hep-th",
    doi = "10.1007/JHEP07(2014)152",
    journal = "JHEP",
    volume = "07",
    pages = "152",
    year = "2014"
}

@article{He:2014laa,
    author = "He, Temple and Lysov, Vyacheslav and Mitra, Prahar and Strominger, Andrew",
    title = "{BMS supertranslations and Weinberg\textquoteright{}s soft graviton theorem}",
    eprint = "1401.7026",
    archivePrefix = "arXiv",
    primaryClass = "hep-th",
    doi = "10.1007/JHEP05(2015)151",
    journal = "JHEP",
    volume = "05",
    pages = "151",
    year = "2015"
}

@article{Cachazo:2014fwa,
    author = "Cachazo, Freddy and Strominger, Andrew",
    title = "{Evidence for a New Soft Graviton Theorem}",
    eprint = "1404.4091",
    archivePrefix = "arXiv",
    primaryClass = "hep-th",
    month = "4",
    year = "2014"
}

@article{Kapec:2014opa,
    author = "Kapec, Daniel and Lysov, Vyacheslav and Pasterski, Sabrina and Strominger, Andrew",
    title = "{Semiclassical Virasoro symmetry of the quantum gravity $ \mathcal{S}$-matrix}",
    eprint = "1406.3312",
    archivePrefix = "arXiv",
    primaryClass = "hep-th",
    doi = "10.1007/JHEP08(2014)058",
    journal = "JHEP",
    volume = "08",
    pages = "058",
    year = "2014"
}

@article{Lysov:2014csa,
    author = "Lysov, Vyacheslav and Pasterski, Sabrina and Strominger, Andrew",
    title = "{Low\textquoteright{}s Subleading Soft Theorem as a Symmetry of QED}",
    eprint = "1407.3814",
    archivePrefix = "arXiv",
    primaryClass = "hep-th",
    doi = "10.1103/PhysRevLett.113.111601",
    journal = "Phys. Rev. Lett.",
    volume = "113",
    number = "11",
    pages = "111601",
    year = "2014"
}

@article{He:2014cra,
    author = "He, Temple and Mitra, Prahar and Porfyriadis, Achilleas P. and Strominger, Andrew",
    title = "{New Symmetries of Massless QED}",
    eprint = "1407.3789",
    archivePrefix = "arXiv",
    primaryClass = "hep-th",
    doi = "10.1007/JHEP10(2014)112",
    journal = "JHEP",
    volume = "10",
    pages = "112",
    year = "2014"
}

@article{Kapec:2014zla,
    author = "Kapec, Daniel and Lysov, Vyacheslav and Strominger, Andrew",
    title = "{Asymptotic Symmetries of Massless QED in Even Dimensions}",
    eprint = "1412.2763",
    archivePrefix = "arXiv",
    primaryClass = "hep-th",
    doi = "10.4310/ATMP.2017.v21.n7.a6",
    journal = "Adv. Theor. Math. Phys.",
    volume = "21",
    pages = "1747--1767",
    year = "2017"
}

@article{Kapec:2015vwa,
    author = "Kapec, Daniel and Lysov, Vyacheslav and Pasterski, Sabrina and Strominger, Andrew",
    title = "{Higher-dimensional supertranslations and Weinberg\textquoteright{}s soft graviton theorem}",
    eprint = "1502.07644",
    archivePrefix = "arXiv",
    primaryClass = "gr-qc",
    reportNumber = "CALT-TH-2015-006",
    doi = "10.4310/AMSA.2017.v2.n1.a2",
    journal = "Ann. Math. Sci. Appl.",
    volume = "02",
    pages = "69--94",
    year = "2017"
}

@article{He:2015zea,
    author = "He, Temple and Mitra, Prahar and Strominger, Andrew",
    title = "{2D Kac-Moody Symmetry of 4D Yang-Mills Theory}",
    eprint = "1503.02663",
    archivePrefix = "arXiv",
    primaryClass = "hep-th",
    doi = "10.1007/JHEP10(2016)137",
    journal = "JHEP",
    volume = "10",
    pages = "137",
    year = "2016"
}

@article{Kapec:2015ena,
    author = "Kapec, Daniel and Pate, Monica and Strominger, Andrew",
    title = "{New Symmetries of QED}",
    eprint = "1506.02906",
    archivePrefix = "arXiv",
    primaryClass = "hep-th",
    doi = "10.4310/ATMP.2017.v21.n7.a7",
    journal = "Adv. Theor. Math. Phys.",
    volume = "21",
    pages = "1769--1785",
    year = "2017"
}

@article{Strominger:2015bla,
    author = "Strominger, Andrew",
    title = "{Magnetic Corrections to the Soft Photon Theorem}",
    eprint = "1509.00543",
    archivePrefix = "arXiv",
    primaryClass = "hep-th",
    doi = "10.1103/PhysRevLett.116.031602",
    journal = "Phys. Rev. Lett.",
    volume = "116",
    number = "3",
    pages = "031602",
    year = "2016"
}

@article{Dumitrescu:2015fej,
    author = "Dumitrescu, Thomas T. and He, Temple and Mitra, Prahar and Strominger, Andrew",
    title = "{Infinite-dimensional fermionic symmetry in supersymmetric gauge theories}",
    eprint = "1511.07429",
    archivePrefix = "arXiv",
    primaryClass = "hep-th",
    doi = "10.1007/JHEP08(2021)051",
    journal = "JHEP",
    volume = "08",
    pages = "051",
    year = "2021"
}

@article{Campiglia:2014yka,
    author = "Campiglia, Miguel and Laddha, Alok",
    title = "{Asymptotic symmetries and subleading soft graviton theorem}",
    eprint = "1408.2228",
    archivePrefix = "arXiv",
    primaryClass = "hep-th",
    doi = "10.1103/PhysRevD.90.124028",
    journal = "Phys. Rev. D",
    volume = "90",
    number = "12",
    pages = "124028",
    year = "2014"
}

@article{Liu:2014vva,
    author = "Liu, Zheng-Wen",
    title = "{Soft theorems in maximally supersymmetric theories}",
    eprint = "1410.1616",
    archivePrefix = "arXiv",
    primaryClass = "hep-th",
    doi = "10.1140/epjc/s10052-015-3304-1",
    journal = "Eur. Phys. J. C",
    volume = "75",
    number = "3",
    pages = "105",
    year = "2015"
}

@article{Larkoski:2014bxa,
    author = "Larkoski, Andrew J. and Neill, Duff and Stewart, Iain W.",
    title = "{Soft Theorems from Effective Field Theory}",
    eprint = "1412.3108",
    archivePrefix = "arXiv",
    primaryClass = "hep-th",
    reportNumber = "MIT-CTP-4568",
    doi = "10.1007/JHEP06(2015)077",
    journal = "JHEP",
    volume = "06",
    pages = "077",
    year = "2015"
}

@article{Campiglia:2015qka,
    author = "Campiglia, Miguel and Laddha, Alok",
    title = "{Asymptotic symmetries of QED and Weinberg\textquoteright{}s soft photon theorem}",
    eprint = "1505.05346",
    archivePrefix = "arXiv",
    primaryClass = "hep-th",
    doi = "10.1007/JHEP07(2015)115",
    journal = "JHEP",
    volume = "07",
    pages = "115",
    year = "2015"
}

@article{Campiglia:2015kxa,
    author = "Campiglia, Miguel and Laddha, Alok",
    title = "{Asymptotic symmetries of gravity and soft theorems for massive particles}",
    eprint = "1509.01406",
    archivePrefix = "arXiv",
    primaryClass = "hep-th",
    doi = "10.1007/JHEP12(2015)094",
    journal = "JHEP",
    volume = "12",
    pages = "094",
    year = "2015"
}

@article{Campiglia:2016hvg,
    author = "Campiglia, Miguel and Laddha, Alok",
    title = "{Subleading soft photons and large gauge transformations}",
    eprint = "1605.09677",
    archivePrefix = "arXiv",
    primaryClass = "hep-th",
    doi = "10.1007/JHEP11(2016)012",
    journal = "JHEP",
    volume = "11",
    pages = "012",
    year = "2016"
}

@article{Hirai:2018ijc,
    author = "Hirai, Hayato and Sugishita, Sotaro",
    title = "{Conservation Laws from Asymptotic Symmetry and Subleading Charges in QED}",
    eprint = "1805.05651",
    archivePrefix = "arXiv",
    primaryClass = "hep-th",
    reportNumber = "OU-HET-971",
    doi = "10.1007/JHEP07(2018)122",
    journal = "JHEP",
    volume = "07",
    pages = "122",
    year = "2018"
}

@article{Campiglia:2018see,
    author = "Campiglia, Miguel and Freidel, Laurent and Hopfmueller, Florian and Soni, Ronak M.",
    title = "{Scalar Asymptotic Charges and Dual Large Gauge Transformations}",
    eprint = "1810.04213",
    archivePrefix = "arXiv",
    primaryClass = "hep-th",
    reportNumber = "TIFR/TH/18-22",
    doi = "10.1007/JHEP04(2019)003",
    journal = "JHEP",
    volume = "04",
    pages = "003",
    year = "2019"
}

@article{Francia:2018jtb,
    author = "Francia, Dario and Heissenberg, Carlo",
    title = "{Two-Form Asymptotic Symmetries and Scalar Soft Theorems}",
    eprint = "1810.05634",
    archivePrefix = "arXiv",
    primaryClass = "hep-th",
    doi = "10.1103/PhysRevD.98.105003",
    journal = "Phys. Rev. D",
    volume = "98",
    number = "10",
    pages = "105003",
    year = "2018"
}

@article{Campiglia:2021oqz,
    author = "Campiglia, Miguel and Peraza, Javier",
    title = "{Charge algebra for non-abelian large gauge symmetries at O(r)}",
    eprint = "2111.00973",
    archivePrefix = "arXiv",
    primaryClass = "hep-th",
    doi = "10.1007/JHEP12(2021)058",
    journal = "JHEP",
    volume = "12",
    pages = "058",
    year = "2021"
}

@article{Nagy:2022xxs,
    author = "Nagy, Silvia and Peraza, Javier",
    title = "{Radiative phase space extensions at all orders in r for self-dual Yang-Mills and gravity}",
    eprint = "2211.12991",
    archivePrefix = "arXiv",
    primaryClass = "hep-th",
    doi = "10.1007/JHEP02(2023)202",
    journal = "JHEP",
    volume = "02",
    pages = "202",
    year = "2023"
}

@article{Peraza:2023ivy,
    author = "Peraza, Javier",
    title = "{Renormalized electric and magnetic charges for O(r$^{n}$) large gauge symmetries}",
    eprint = "2301.05671",
    archivePrefix = "arXiv",
    primaryClass = "hep-th",
    doi = "10.1007/JHEP01(2024)175",
    journal = "JHEP",
    volume = "01",
    pages = "175",
    year = "2024"
}

@article{Nagy:2024jua,
    author = "Nagy, Silvia and Peraza, Javier and Pizzolo, Giorgio",
    title = "{Infinite-dimensional hierarchy of recursive extensions for all sub$^{n}$-leading soft effects in Yang-Mills}",
    eprint = "2407.13556",
    archivePrefix = "arXiv",
    primaryClass = "hep-th",
    doi = "10.1007/JHEP12(2024)068",
    journal = "JHEP",
    volume = "12",
    pages = "068",
    year = "2024"
}

@article{Strominger:2014pwa,
    author = "Strominger, Andrew and Zhiboedov, Alexander",
    title = "{Gravitational Memory, BMS Supertranslations and Soft Theorems}",
    eprint = "1411.5745",
    archivePrefix = "arXiv",
    primaryClass = "hep-th",
    doi = "10.1007/JHEP01(2016)086",
    journal = "JHEP",
    volume = "01",
    pages = "086",
    year = "2016"
}

@article{Pasterski:2015tva,
    author = "Pasterski, Sabrina and Strominger, Andrew and Zhiboedov, Alexander",
    title = "{New Gravitational Memories}",
    eprint = "1502.06120",
    archivePrefix = "arXiv",
    primaryClass = "hep-th",
    doi = "10.1007/JHEP12(2016)053",
    journal = "JHEP",
    volume = "12",
    pages = "053",
    year = "2016"
}

@article{Pate:2017fgt,
    author = "Pate, Monica and Raclariu, Ana-Maria and Strominger, Andrew",
    title = "{Gravitational Memory in Higher Dimensions}",
    eprint = "1712.01204",
    archivePrefix = "arXiv",
    primaryClass = "hep-th",
    doi = "10.1007/JHEP06(2018)138",
    journal = "JHEP",
    volume = "06",
    pages = "138",
    year = "2018"
}

@article{Hamada:2017atr,
    author = "Hamada, Yuta and Sugishita, Sotaro",
    title = "{Soft pion theorem, asymptotic symmetry and new memory effect}",
    eprint = "1709.05018",
    archivePrefix = "arXiv",
    primaryClass = "hep-th",
    reportNumber = "OU-HET-944",
    doi = "10.1007/JHEP11(2017)203",
    journal = "JHEP",
    volume = "11",
    pages = "203",
    year = "2017"
}

@article{Hamada:2018cjj,
    author = "Hamada, Yuta and Sugishita, Sotaro",
    title = "{Notes on the gravitational, electromagnetic and axion memory effects}",
    eprint = "1803.00738",
    archivePrefix = "arXiv",
    primaryClass = "hep-th",
    reportNumber = "OU-HET-962",
    doi = "10.1007/JHEP07(2018)017",
    journal = "JHEP",
    volume = "07",
    pages = "017",
    year = "2018"
}

@article{Campiglia:2017dpg,
    author = "Campiglia, Miguel and Coito, Leonardo and Mizera, Sebastian",
    title = "{Can scalars have asymptotic symmetries?}",
    eprint = "1703.07885",
    archivePrefix = "arXiv",
    primaryClass = "hep-th",
    doi = "10.1103/PhysRevD.97.046002",
    journal = "Phys. Rev. D",
    volume = "97",
    number = "4",
    pages = "046002",
    year = "2018"
}

@article{Henneaux:2018mgn,
    author = "Henneaux, Marc and Troessaert, C\'edric",
    title = "{Asymptotic structure of a massless scalar field and its dual two-form field at spatial infinity}",
    eprint = "1812.07445",
    archivePrefix = "arXiv",
    primaryClass = "hep-th",
    doi = "10.1007/JHEP05(2019)147",
    journal = "JHEP",
    volume = "05",
    pages = "147",
    year = "2019"
}

@article{Kapec:2017tkm,
    author = "Kapec, Daniel and Perry, Malcolm and Raclariu, Ana-Maria and Strominger, Andrew",
    title = "{Infrared Divergences in QED, Revisited}",
    eprint = "1705.04311",
    archivePrefix = "arXiv",
    primaryClass = "hep-th",
    doi = "10.1103/PhysRevD.96.085002",
    journal = "Phys. Rev. D",
    volume = "96",
    number = "8",
    pages = "085002",
    year = "2017"
}

@article{Gabai:2016kuf,
    author = "Gabai, Barak and Sever, Amit",
    title = "{Large gauge symmetries and asymptotic states in QED}",
    eprint = "1607.08599",
    archivePrefix = "arXiv",
    primaryClass = "hep-th",
    doi = "10.1007/JHEP12(2016)095",
    journal = "JHEP",
    volume = "12",
    pages = "095",
    year = "2016"
}

@article{Hirai:2019gio,
    author = "Hirai, Hayato and Sugishita, Sotaro",
    title = "{Dressed states from gauge invariance}",
    eprint = "1901.09935",
    archivePrefix = "arXiv",
    primaryClass = "hep-th",
    reportNumber = "OU-HET-998",
    doi = "10.1007/JHEP06(2019)023",
    journal = "JHEP",
    volume = "06",
    pages = "023",
    year = "2019"
}

@article{Hirai:2020kzx,
    author = "Hirai, Hayato and Sugishita, Sotaro",
    title = "{IR finite S-matrix by gauge invariant dressed states}",
    eprint = "2009.11716",
    archivePrefix = "arXiv",
    primaryClass = "hep-th",
    reportNumber = "NIT-KMP-202001, KEK-TH-2261",
    doi = "10.1007/JHEP02(2021)025",
    journal = "JHEP",
    volume = "02",
    pages = "025",
    year = "2021"
}

@article{Hirai:2022yqw,
    author = "Hirai, Hayato and Sugishita, Sotaro",
    title = "{Dress code for infrared safe scattering in QED}",
    eprint = "2209.00608",
    archivePrefix = "arXiv",
    primaryClass = "hep-th",
    doi = "10.1093/ptep/ptad057",
    journal = "PTEP",
    volume = "2023",
    number = "5",
    pages = "053B04",
    year = "2023"
}

@article{Mirbabayi:2016axw,
    author = "Mirbabayi, Mehrdad and Porrati, Massimo",
    title = "{Dressed Hard States and Black Hole Soft Hair}",
    eprint = "1607.03120",
    archivePrefix = "arXiv",
    primaryClass = "hep-th",
    doi = "10.1103/PhysRevLett.117.211301",
    journal = "Phys. Rev. Lett.",
    volume = "117",
    number = "21",
    pages = "211301",
    year = "2016"
}

@article{Carney:2018ygh,
    author = "Carney, Daniel and Chaurette, Laurent and Neuenfeld, Dominik and Semenoff, Gordon",
    title = "{On the need for soft dressing}",
    eprint = "1803.02370",
    archivePrefix = "arXiv",
    primaryClass = "hep-th",
    doi = "10.1007/JHEP09(2018)121",
    journal = "JHEP",
    volume = "09",
    pages = "121",
    year = "2018"
}

@article{Neuenfeld:2018fdw,
    author = "Neuenfeld, Dominik",
    title = "{Infrared-safe scattering without photon vacuum transitions and time-dependent decoherence}",
    eprint = "1810.11477",
    archivePrefix = "arXiv",
    primaryClass = "hep-th",
    doi = "10.1007/JHEP11(2021)189",
    journal = "JHEP",
    volume = "11",
    pages = "189",
    year = "2021"
}

@article{Gonzo:2019fai,
    author = "Gonzo, Riccardo and Mc Loughlin, Tristan and Medrano, Diego and Spiering, Anne",
    title = "{Asymptotic charges and coherent states in QCD}",
    eprint = "1906.11763",
    archivePrefix = "arXiv",
    primaryClass = "hep-th",
    reportNumber = "TCD MATH 19-09, SAGEX-19-14",
    doi = "10.1103/PhysRevD.104.025019",
    journal = "Phys. Rev. D",
    volume = "104",
    number = "2",
    pages = "025019",
    year = "2021"
}

@article{Choi:2019rlz,
    author = "Choi, Sangmin and Akhoury, Ratindranath",
    title = "{Subleading soft dressings of asymptotic states in QED and perturbative quantum gravity}",
    eprint = "1907.05438",
    archivePrefix = "arXiv",
    primaryClass = "hep-th",
    doi = "10.1007/JHEP09(2019)031",
    journal = "JHEP",
    volume = "09",
    pages = "031",
    year = "2019"
}

@article{Oertel:2026wsm,
    author = "Oertel, Brett and Moult, Ian and Pasterski, Sabrina",
    title = "{Asymptotic charges as detectors and the memory effect in massive QED and perturbative quantum gravity}",
    eprint = "2604.19866",
    archivePrefix = "arXiv",
    primaryClass = "hep-th",
    month = "4",
    year = "2026"
}

@article{Oertel:2026oqv,
    author = "Oertel, Brett",
    title = "{Finite-time memory detectors and fully constraining Faddeev-Kulish dressings in QED and gravity}",
    eprint = "2605.06774",
    archivePrefix = "arXiv",
    primaryClass = "hep-th",
    month = "5",
    year = "2026"
}

@article{Donnelly:2016auv,
    author = "Donnelly, William and Freidel, Laurent",
    title = "{Local subsystems in gauge theory and gravity}",
    eprint = "1601.04744",
    archivePrefix = "arXiv",
    primaryClass = "hep-th",
    doi = "10.1007/JHEP09(2016)102",
    journal = "JHEP",
    volume = "09",
    pages = "102",
    year = "2016"
}

@article{Carrozza:2021gju,
    author = "Carrozza, Sylvain and Hoehn, Philipp A.",
    title = "{Edge modes as reference frames and boundary actions from post-selection}",
    eprint = "2109.06184",
    archivePrefix = "arXiv",
    primaryClass = "hep-th",
    doi = "10.1007/JHEP02(2022)172",
    journal = "JHEP",
    volume = "02",
    pages = "172",
    year = "2022"
}

@article{Araujo-Regado:2024dpr,
    author = "Araujo-Regado, Goncalo and Hoehn, Philipp A. and Sartini, Francesco and Tomova, Bilyana",
    title = "{Soft edges: the many links between soft and edge modes}",
    eprint = "2412.14548",
    archivePrefix = "arXiv",
    primaryClass = "hep-th",
    doi = "10.1007/JHEP07(2025)180",
    journal = "JHEP",
    volume = "07",
    pages = "180",
    year = "2025"
}

@article{Ball:2024hqe,
    author = "Ball, Adam and Law, Y. T. Albert and Wong, Gabriel",
    title = "{Dynamical edge modes and entanglement in Maxwell theory}",
    eprint = "2403.14542",
    archivePrefix = "arXiv",
    primaryClass = "hep-th",
    doi = "10.1007/JHEP09(2024)032",
    journal = "JHEP",
    volume = "09",
    pages = "032",
    year = "2024"
}

@article{Hoehn:2025pmx,
    author = "Hoehn, Philipp A. and Kirklin, Josh",
    title = "{Fighting non-locality with non-locality: microcausality and boundary conditions in QED}",
    eprint = "2512.16898",
    archivePrefix = "arXiv",
    primaryClass = "hep-th",
    month = "12",
    year = "2025"
}

@article{Shimizu:2026zug,
    author = "Shimizu, Keito and Sugishita, Sotaro",
    title = "{Revisiting boundary electromagnetic duality and edge modes}",
    eprint = "2605.27870",
    archivePrefix = "arXiv",
    primaryClass = "hep-th",
    reportNumber = "KUNS-3105, EPHOU-26-05, RIKEN-iTHEMS-Report-26",
    month = "5",
    year = "2026"
}

@article{PhysRevLett.64.216,
  title = {Edge channels for the fractional quantum Hall effect},
  author = {Beenakker, C. W. J.},
  journal = {Phys. Rev. Lett.},
  volume = {64},
  issue = {2},
  pages = {216--219},
  numpages = {0},
  year = {1990},
  month = {Jan},
  publisher = {American Physical Society},
  doi = {10.1103/PhysRevLett.64.216},
  url = {https://link.aps.org/doi/10.1103/PhysRevLett.64.216}
}

@article{MacDonald:1990zz,
    author = "MacDonald, A. H.",
    title = "{Edge states in the fractional-quantum-Hall-effect regime}",
    doi = "10.1103/PhysRevLett.64.220",
    journal = "Phys. Rev. Lett.",
    volume = "64",
    pages = "220--223",
    year = "1990"
}

@article{Wen:1990se,
    author = "Wen, X. G.",
    title = "{Chiral Luttinger Liquid and the Edge Excitations in the Fractional Quantum Hall States}",
    reportNumber = "IASSNS-HEP-90-14",
    doi = "10.1103/PhysRevB.41.12838",
    journal = "Phys. Rev. B",
    volume = "41",
    pages = "12838--12844",
    year = "1990"
}

@article{Wen:1995qn,
    author = "Wen, Xiao-Gang",
    title = "{Topological orders and edge excitations in FQH states}",
    eprint = "cond-mat/9506066",
    archivePrefix = "arXiv",
    reportNumber = "PRINT-95-148 (MIT)",
    doi = "10.1080/00018739500101566",
    journal = "Adv. Phys.",
    volume = "44",
    number = "5",
    pages = "405--473",
    year = "1995"
}

@article{Carlip:1994gy,
    author = "Carlip, Steven",
    title = "{The Statistical mechanics of the (2+1)-dimensional black hole}",
    eprint = "gr-qc/9409052",
    archivePrefix = "arXiv",
    reportNumber = "UCD-94-32, NI-94011",
    doi = "10.1103/PhysRevD.51.632",
    journal = "Phys. Rev. D",
    volume = "51",
    pages = "632--637",
    year = "1995"
}

@article{Ashtekar:1997yu,
    author = "Ashtekar, A. and Baez, J. and Corichi, A. and Krasnov, Kirill",
    title = "{Quantum geometry and black hole entropy}",
    eprint = "gr-qc/9710007",
    archivePrefix = "arXiv",
    reportNumber = "CGPG-97-9-3",
    doi = "10.1103/PhysRevLett.80.904",
    journal = "Phys. Rev. Lett.",
    volume = "80",
    pages = "904--907",
    year = "1998"
}

@inproceedings{Balachandran:1995qa,
    author = "Balachandran, A. P. and Chandar, L. and Momen, Arshad",
    title = "{Edge states in canonical gravity}",
    booktitle = "{17th Annual MRST (Montreal-Rochester-Syracuse-Toronto) Meeting on High-energy Physics}",
    eprint = "gr-qc/9506006",
    archivePrefix = "arXiv",
    reportNumber = "SU-4240-610",
    month = "5",
    year = "1995"
}

@article{Carlip:2005zn,
    author = "Carlip, Steven",
    title = "{Conformal field theory, (2+1)-dimensional gravity, and the BTZ black hole}",
    eprint = "gr-qc/0503022",
    archivePrefix = "arXiv",
    reportNumber = "UCD-05-02",
    doi = "10.1088/0264-9381/22/12/R01",
    journal = "Class. Quant. Grav.",
    volume = "22",
    pages = "R85--R124",
    year = "2005"
}

@article{Shimizu:2025hfl,
    author = "Shimizu, Keito and Sugishita, Sotaro",
    title = "{Asymptotic symmetry and confinement in three-dimensional QED}",
    eprint = "2503.20173",
    archivePrefix = "arXiv",
    primaryClass = "hep-th",
    reportNumber = "KUNS-3041",
    doi = "10.1103/x92g-9w2h",
    journal = "Phys. Rev. D",
    volume = "112",
    number = "10",
    pages = "105001",
    year = "2025"
}

@article{Freidel:2018fsk,
    author = "Freidel, Laurent and Pranzetti, Daniele",
    title = "{Electromagnetic duality and central charge}",
    eprint = "1806.03161",
    archivePrefix = "arXiv",
    primaryClass = "hep-th",
    doi = "10.1103/PhysRevD.98.116008",
    journal = "Phys. Rev. D",
    volume = "98",
    number = "11",
    pages = "116008",
    year = "2018"
}

@article{Hosseinzadeh:2018dkh,
    author = "Hosseinzadeh, V. and Seraj, A. and Sheikh-Jabbari, M. M.",
    title = "{Soft Charges and Electric-Magnetic Duality}",
    eprint = "1806.01901",
    archivePrefix = "arXiv",
    primaryClass = "hep-th",
    reportNumber = "IPM-P-2018-030",
    doi = "10.1007/JHEP08(2018)102",
    journal = "JHEP",
    volume = "08",
    pages = "102",
    year = "2018"
}

@article{Choi:2019sjs,
    author = "Choi, Sangmin and Akhoury, Ratindranath",
    title = "{Magnetic soft charges, dual supertranslations, and {\textquoteright}t Hooft line dressings}",
    eprint = "1912.02224",
    archivePrefix = "arXiv",
    primaryClass = "hep-th",
    doi = "10.1103/PhysRevD.102.025001",
    journal = "Phys. Rev. D",
    volume = "102",
    number = "2",
    pages = "025001",
    year = "2020"
}

@article{Mathieu:2020fwg,
    author = "Mathieu, Philippe and Teh, Nicholas J.",
    title = "{Boundary electromagnetic duality from homological edge modes}",
    eprint = "2102.06799",
    archivePrefix = "arXiv",
    primaryClass = "math-ph",
    doi = "10.1007/JHEP07(2021)192",
    journal = "JHEP",
    volume = "21",
    pages = "192",
    year = "2020"
}

@article{Henneaux:2020nxi,
    author = "Henneaux, Marc and Troessaert, C{\'e}dric",
    title = "{A note on electric-magnetic duality and soft charges}",
    eprint = "2004.05668",
    archivePrefix = "arXiv",
    primaryClass = "hep-th",
    doi = "10.1007/JHEP06(2020)081",
    journal = "JHEP",
    volume = "06",
    pages = "081",
    year = "2020"
}

@article{Geiller:2021gdk,
    author = "Geiller, Marc and Jai-akson, Puttarak and Osumanu, Abdulmajid and Pranzetti, Daniele",
    title = "{Electromagnetic duality and central charge from first order formulation}",
    eprint = "2107.05443",
    archivePrefix = "arXiv",
    primaryClass = "hep-th",
    journal = "J. Part. Phys. Cosmol.",
    volume = "1",
    pages = "4",
    year = "2026"
}

@article{Witten:1979ey,
    author = "Witten, Edward",
    title = "{Dyons of Charge e theta/2 pi}",
    reportNumber = "CERN-TH-2724",
    doi = "10.1016/0370-2693(79)90838-4",
    journal = "Phys. Lett. B",
    volume = "86",
    pages = "283--287",
    year = "1979"
}

@article{Harlow:2019yfa,
    author = "Harlow, Daniel and Wu, Jie-Qiang",
    title = "{Covariant phase space with boundaries}",
    eprint = "1906.08616",
    archivePrefix = "arXiv",
    primaryClass = "hep-th",
    doi = "10.1007/JHEP10(2020)146",
    journal = "JHEP",
    volume = "10",
    pages = "146",
    year = "2020"
}
\bibliographystyle{JHEP.bst}

\end{document}